\documentstyle[aps,multicol,pra,epsfig]{revtex}
\def\bea{\begin{eqnarray}}
\def\eea{\end{eqnarray}}
\def\be{\begin{equation}}
\def\ee{\end{equation}}
\begin{document}
\draft
\author{Zbyszek P. Karkuszewski$^{1,2}$, Krzysztof Sacha$^{2}$, 
and  Augusto Smerzi$^{1}$}
\address{
 $^1$Theoretical Division and Center for Nonlinear Studies,
        Los Alamos National Laboratory,
        Los Alamos, NM 87545 USA,\\
$^2$Instytut Fizyki,
  Uniwersytet Jagiello\'nski,
 ulica Reymonta 4, PL-30-059 Krak\'ow, Poland \\
}
\title{Mean field
loops versus quantum anti-crossing nets \\
in trapped Bose-Einstein condensates
}
\date{\today}
\maketitle
\begin{abstract}
We study a Bose-Einstein condensate trapped in an asymmetric double well 
potential. Solutions of the time-independent Gross-Pitaevskii equation 
reveal intrinsic loops in the energy (or chemical potential) 
level behavior when the shape of the potential is varied. We investigate
the corresponding behavior of the quantum (many-body) energy levels. 
Applying the two-mode approximation to the bosonic field operators, we show 
that the quantum energy levels create an anti-crossing 
net inside the region bounded by the loop of the mean field solution.
\end{abstract}
\pacs{PACS: 03.75.Fi,05.30.Jp,32.80.Pj}
\begin{multicols}{2}

Experimental realizations of a Bose-Einstein condensate
(BEC) in cooled and trapped atomic gases \cite{bec}
are triggering many efforts to explore new quantum phenomena
on a macroscopic level. The weak interaction between the condensate
atoms allows for a precise tailoring
that is not feasible experimentally with 
other superfluid systems such as the $^4$He-II. 
In particular, interference phenomena have been investigated
in the case of a condensate trapped in both
double \cite{ketterle97} and multiple \cite{kasevich}
wells potentials, and the possibility of tunneling has resulted in an
oscillatory (Josephson) atomic flow across adjacent traps \cite{cataliotti01}. 
A theoretical description of the process carried out within 
the mean field (Gross-Pitaevskii equation (GPE) \cite{dalfovo99}) 
approximation predicts a rich dynamical phase diagram 
\cite{smerzi}. Quantum corrections show that the 
mean field description remains valid for a very long but finite period of time 
up to the moment when the system is able to detect tiny differences in energy
between the multi-particle levels \cite{walls,raghavan99,holthaus}.

The excitation of solitons and vortices in a condensate 
by an adiabatic deformation of the trapping potential has been 
recently investigated in \cite{kark01,dksz}. 
However, we have found that some levels of the time-independent GPE reveal 
loop-like structures when varying a parameter of the potential
\cite{dksz}. 
Solutions of GPE which correspond to eigenvalues on those loops are typical of
nonlinear equations and do not have a linear ($g_0=0$ in (\ref{gpe})) 
single particle counterpart. This class of GPE solutions has been  
investigated in \cite{presilla} for a symmetric double well potential.
The loop-like behavior in the solution of the GPE has been also observed
for Bloch bands spectrum in Ref.~\cite{biao}. There, the appearance of the loops
results in the breakdown of the Bloch oscillations due 
to non-zero adiabatic tunneling
into the upper band.

One may wonder about the quantum (multi-particle) equivalent of such loops. 
In this case, indeed,
no loops are expected because the many-body Hamiltonian 
is a Hermitian operator. 
It has been proved in \cite{lieb} that 
the ground state and the associated energy of GPE for repulsive
atom-atom interactions (positive scattering length $a_s$) are asymptotically 
exact when the number of particles $N \to \infty$ while $Na_s$ is kept constant.
To the best of our knowledge there is no similar theorem concerning 
the  negative $a_s$ case and the
excited eigenstates of the GPE. Therefore, 
it is important to understand how those mean-field loops are approached  
by the eigenenergies of the quantum Hamiltonian which, we remark,
cannot form any loop.

The aim of the present paper is to investigate quantum multiparticle energy
levels of an asymmetric double well potential in the regime where the GPE 
reveals loop-like structures. 
The structures can be observed even for a very weak
interaction between particles, so we use
a two-mode 
approximation (TMA) \cite{walls,smerzi,raghavan99,holthaus} in the quantum
calculations. The results so obtained are compared with the GPE predictions, 
calculated both in the TMA and  
with a numerical integration in a 1D geometry. 

We consider a condensate trapped in a harmonic cigar-like potential 
with an additional Gaussian
perturbation. Freezing the transversal degrees of freedom tightly confining 
the condensate in the same direction, the potential
(in units of a harmonic oscillator corresponding to the longitudinal direction)
reads
\be
V(x)=\frac{x^2}{2}+U_0\arctan(x_0)\exp\left(-\frac{(x-x_0)^2}{2\sigma^2}\right).
\label{pot}
\ee
Within a certain window of parameters, the potential (\ref{pot}) has a double
well shape \cite{kark01,dksz}. 
The many-body Hamiltonian governing the full $N$-particle problem
\cite{foot1} is,
\be
\hat{H}=\int dx\left[\hat{\psi}^\dagger H_0\hat{\psi}+\frac{g_0}{2}
\hat{\psi}^\dagger\hat{\psi}^\dagger\hat{\psi}\hat{\psi}\right],
\ee 
where $H_0=-\frac{1}{2}\frac{d^2}{dx^2}+V(x)$. $\hat{H}$
can be approximated expanding the field operators in terms of the
local modes, $\hat{\psi}\approx u_1(x)\hat{c}_1+u_2(x)\hat{c}_2$.
$u_j(x)$
are localized solutions of the Schr\"odinger equation solved
for each well of the potential 
separately (see below). This, provided that
$\int u_1^*(x)u_2(x)dx\approx 0$, yields 
\bea
\hat{H}&\approx&E_1\hat{c}_1^\dagger\hat{c}_1+E_2\hat{c}_2^\dagger\hat{c}_2
+\frac{\Omega}{2}(\hat{c}_2^\dagger\hat{c}_1+\hat{c}_1^\dagger\hat{c}_2) \cr
&&+\frac{g_0}{2V_1}(\hat{c}_1^\dagger)^2(\hat{c}_1)^2+
\frac{g_0}{2V_2}(\hat{c}_2^\dagger)^2(\hat{c}_2)^2,
\label{2mod}
\eea
where $E_j=\int u_j^*(x)H_0u_j(x)dx$, $\Omega=2\int u_1^*(x)H_0u_2(x)dx$
and the effective mode volume
of each well
$V_j^{-1}=\int |u_j(x)|^4dx$.

\begin{figure}
\centering
{\epsfig{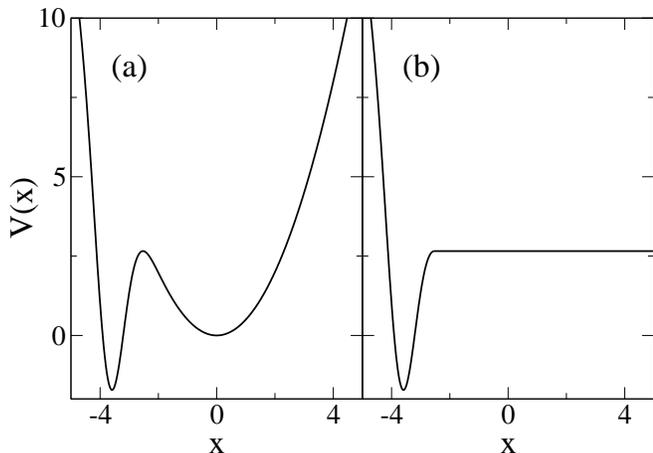}}
\caption{
(a): plot of the potential ({\protect \ref{pot}}) for $U_0=6.4$,
$\sigma=0.5$ and $x_0=-3.7$. (b): single well potential obtained by taking
the left part of the potential plotted in panel (a) up to the point
corresponding to the top of the barrier and in the remaining space the potential
is kept constant. 
The lowest eigenstate of such a single well potential constitutes 
the mode $u_1(x)$. In an analogous way the mode $u_2(x)$ of the right potential 
well is defined.
}
\label{singpot}
\end{figure}

\begin{figure}
\centering
{\epsfig{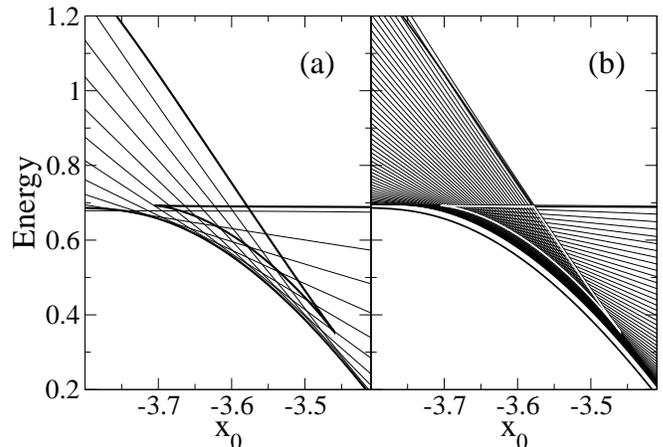}}
\caption{
Energy levels of a condensate per particle as a function of $x_0$, for $Ng_0=1$. 
The other potential parameters are: $U_0=6.4$ and $\sigma=0.5$. 
Thick lines, in both panels, correspond to the solutions of the
GPE ({\protect \ref{gpe}}), i.e. $E/N=\mu-(Ng_0/2)\int |\varphi(x)|^4dx$, 
while lines are the results of the quantum
two-mode approximation ({\protect \ref{2mod}}) for number of particles 
$N=10$ (a) and $N=50$ (b).
}
\label{gplus}
\end{figure}

\begin{figure}
\centering
{\epsfig{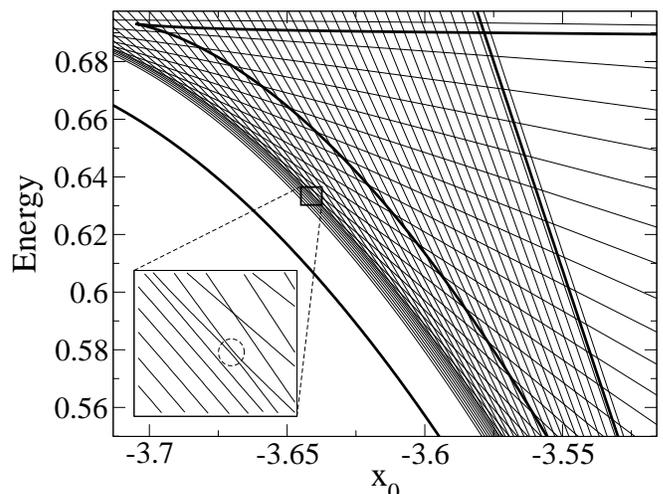}}
\caption{
Magnified anti-crossings structure of Fig.\ref{gplus}b. Anti-crossings are so
tiny that they become visible in the insert (in the circle).
}
\label{gplusmag}
\end{figure}

\begin{figure}
\centering
{\epsfig{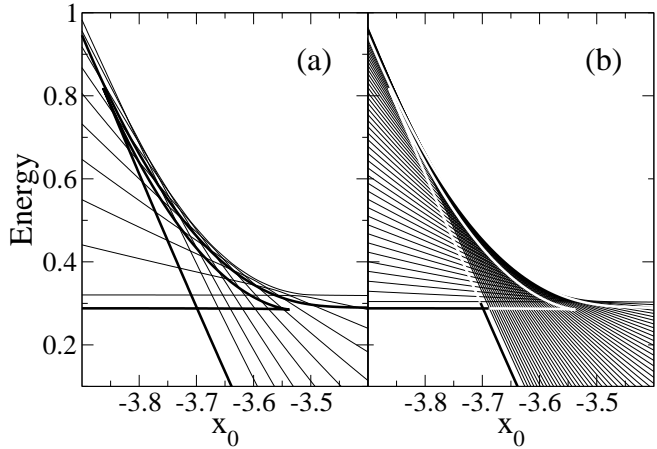}}
\caption{
The same as in Fig.~{\protect \ref{gplus}} but for $Ng_0=-1$.
}
\label{gminus}
\end{figure}

The Bose operators $\hat{c}_j$, that annihilate a particle at the respective
well of the potential, obey the usual commutation relation
$[\hat{c}_j,\hat{c}_k^\dagger]=\delta_{jk}$.
To define the modes $u_j(x)$, we have separated each well from the potential 
(\ref{pot}) as shown in Fig.~\ref{singpot} \cite{foot2}. Given the modes
one can immediately calculate the parameters of the Hamiltonian (\ref{2mod}).
The solutions of the time-independent GPE,
\be
H_0\varphi(x)+Ng_0|\varphi(x)|^2\varphi(x)=\mu\varphi(x),
\label{gpe}
\ee
where $|\varphi(x)|^2$ is normalized to unity and $N$ denotes the number of 
particles in a BEC, give the chemical potential levels $\mu$ of the condensate 
\cite{dalfovo99}. For a very small
nonlinearity parameter, $Ng_0$, changing the parameter of the potential
$x_0$ (for $U_0=6.4$ and $\sigma=0.5$), 
a narrow avoided crossing between the ground and first excited 
levels occurs around $x_0=-3.6$ \cite{kark01}. 
However, when $Ng_0$ increases the usual
avoided crossing turns into a more 
complicated structure where the excited level 
reveals a loop-like character, see Fig.~\ref{gplus}. To compare this
single-particle description of the system with the quantum multi-particle
picture we have diagonalized the Hamiltonian (\ref{2mod}) for $N=10$ and 
$N=50$
adjusting $g_0$ so that $Ng_0=1$ in both cases.
Figure~\ref{gplusmag} shows 
that quantum energy levels undergo a net of anti-crossings and
the GPE solutions provide an envelope of this anti-crossings structure.
The lowest quantum energy level corresponds to the ground state of the BEC while
the highest one corresponds to all particles occupying the upper level of the
trap. Both states can be calculated by solving the GPE equation. However,
in the middle of the figure not only the lowest and highest energy states 
can be represented by a mean field wavefunction which corresponds to all atoms 
in the same one-particle state. Indeed, the existence of the loop indicates 
that there are two more of such states. 
As expected, also in the attractive case we 
observe a loop-like structure (see Fig.~\ref{gminus}) which, however,
is on the ground 
state energy level.
In the two-mode approximation, the conditions for the formation of loops
can be easily calculated analytically (see below).

\begin{figure}
\centering
{\epsfig{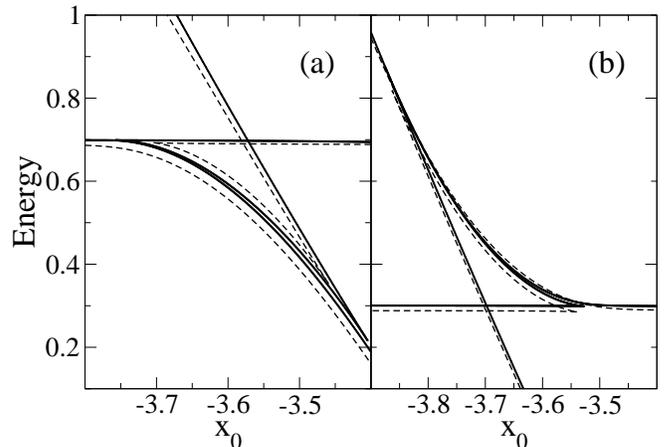}}
\caption{
Comparison of the two-mode approximation applied to the GPE (solid lines) 
with the solutions of the full GPE (dashed lines). Panel (a) corresponds to
$Ng_0=1$, panel (b) to $Ng_0=-1$.
}
\label{2gpe}
\end{figure}

Switching to the quantum two-mode approximation we found that the
energy levels 
reveal (similarly as for the repulsive interaction) an 
anti-crossing structure located within the loop of the mean field 
results (Fig.~\ref{gminus}).
Let us now carry out the two-mode approximation also in the mean field
description. Replacing the ansatz wavefunction 
$\varphi(x)=(c_1u_1(x)+c_2u_2(x))/\sqrt{N}$ in (\ref{gpe}) we get a pair of 
coupled equations,
\begin{equation}
\left\{ { {c_1E_1+c_2\frac{\Omega}{2}+|c_1|^2c_1\frac{g_0}{V_1}=c_1\mu}\atop
{c_2E_2+c_1\frac{\Omega}{2}+|c_2|^2c_2\frac{g_0}{V_2}=c_2\mu}}\right. ,
\label{eqset}
\end{equation}
which (with the additional constraint $|c_1|^2+|c_2|^2=N$)
allow us to calculate the coefficients $c_j$ and the chemical potential.
In the range of the parameter $x_0$
where there is no loop of the GPE, we get only 
two solutions of (\ref{eqset}), while  
in the inner region there are four of them forming a loop.
In Fig.~\ref{2gpe} we compare the classical 
two-mode approximation with the results of the GPE both for repulsive 
and attractive interaction between particles. 
The results of the classical two-mode approximation are in a good 
agreement with the full GPE solutions and, therefore, we expect
that the quantum two-mode will also 
be a good approximation of the full 
quantum multi-particle problem. 
We note that the expression for the energy of the $N$-particle condensate
in the classical two-mode approximation,
\bea
H &=&E_1|c_1|^2+E_2|c_2|^2+\frac{\Omega}{2}
(c_1c_2^*+c_1^*c_2) \cr
&&+\frac{g_0}{2V_1}|c_1|^4+\frac{g_0}{2V_2}|c_2|^4,
\eea
is of the same form as the corresponding quantum Hamiltonian (\ref{2mod}), 
but with $c_j$ being c-numbers.

By increasing the number of atoms in the BEC keeping $Ng_0$ fixed, 
the values of the extremal quantum energy levels reproduce the classical 
two-mode approximation results more and more accurately.

\begin{figure}
\centering
{\epsfig{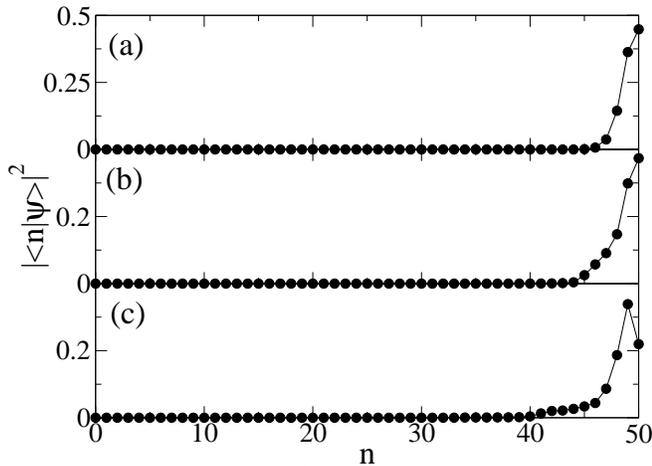}}
\caption{
Overlaps $|\langle n|\psi\rangle|^2$ of the wavefunction 
on the energy eigenstates of the $N=50$ particle system in the harmonic trap
at the end of the laser sweep. Starting with the ground
state of the harmonic trap, i.e. $|\psi(0)\rangle=|n=0\rangle$, we change the
potential ({\protect \ref{pot}}) according to $x_0(t)=x_0(0)+0.05t$ 
with $x_0(0)=-5$. At the end of the sweep the highest energy eigenstates
become populated --- panel (a) corresponds to the noninteracting
particle case (i.e. $g_0=0$), panel (b) to $g_0N=1$ and (c) to $g_0N=-1$.
}
\label{fthree}
\end{figure}

In \cite{kark01} we have suggested that 
starting with a condensate in the ground
state of a harmonic trap and sweeping the trap with a laser beam
(which actually leads to the potential (\ref{pot}) where $x_0$ changes from
some negative value to zero) can be used to
excite a BEC solitary wave by means of a diabatic passage through an avoided
crossing. In the mean field description, the presence of a loop does not break
the process because the ground state wavefunction on one side of such a strange
avoided crossing structure (i.e. a loop) 
is basically the same as the wavefunction of the first excited state on the 
other side so the diabatic scenario applies \cite{dksz}. 
Switching to the quantum description we see that
one avoided crossing of GPE levels, which has to be
passed by each one of the $N$ particles, is just replaced 
by many smaller ones in the multi-particle level structure. 
We expect that when the laser beam is swept with appropriate velocity
it will transfer the atoms from the ground energy 
level on the left hand side of 
Fig.~\ref{gplus} (or Fig.~\ref{gminus}) to the highest energy levels on the
right hand side of Fig.~\ref{gplus} (or Fig.~\ref{gminus}). 
To test such a hypothesis we have carried out quantum time dependent calculations
within the TMA starting with the system in a ground state of the pure harmonic 
trap \cite{tmanew}. In Fig.~\ref{fthree} we show the results of a laser
sweep, i.e. a change of $x_0$ from $-5$ to 0 according to $x_0(t)=-5+0.05t$. 
In the case of the noninteracting particle system ($g_0=0$), if the 
excitation probability of 
a particle from the ground state to the first excited state
in the single particle description is less than 1 (although close to unity),
the corresponding multiparticle picture reveals population of a bunch of 
the eigenstates as shown in Fig.~\ref{fthree}a \cite{kark01}. 
Switching on the interaction 
the effect of the laser sweep is unchanged --- similarly as in the noninteracting
particle case the highest energy eigenstates of the system become populated, see
Fig.~\ref{fthree}b and Fig.~\ref{fthree}c.  
Of course, this is a simplified picture in the two-mode approximation 
that is valid only for a weak interaction between atoms. For stronger 
interactions further investigations are necessary.

In Ref.~\cite{biao} a loop-like behavior was observed in solutions of the
GPE for a BEC in a periodic lattice, implying breakdown of Bloch oscillations
and adiabatic tunneling into the upper band.
The approximation method performed here 
can be easily applied in that case. This allows one to identify
which multi-particle eigenstates are populated in the tunneling process.

To summarize, we have considered the behavior 
of the energy levels of an asymmetric
double well potential when the shape of the potential is changed.
Using the
two-mode approximation in the quantum multi-particle Hamiltonian we have found
that the energy levels undergo a net of avoided crossings. 
Solutions of the Gross-Pitaevskii equation reveal a loop structure which
provides an envelope to the quantum avoided-crossings net. 
We finally note that since the two-mode approximation, applied to the 
mean field description, is 
in a very good agreement with the full numerical solution
of the Gross-Pitaevskii 
equation, the quantum
two-mode approximation can also be considered a good approximation of
the full quantum multi-particle problem at least in the
regime of weak interaction between atoms studied here.

Discussions with Bogdan Damski, Jacek Dziarmaga and Kuba Zakrzewski 
are acknowledged.
KS is grateful to Wojciech H. \.Zurek 
for the hospitality 
during the  preparation of this work. 
Support of KBN under project 5~P03B~088~21 (KS) and of by the U.S. DOE 
are acknowledged.
 
%%%%%references%%%%%%%%%%%%%%%%%%%%%%%%%%%%%%%%%%%%%%%%%%%%%%%%%%%%%%%%%%%%%%

\end{multicols}

\begin{references}
\bibitem{foot1} 
In the harmonic oscillator units, $g_0=g/(\hbar\omega_{||}a_{||})$		
where the coefficient $g=4\pi \hbar^2 a_s/(m\pi a_{\bot}^2)$ 
is renormalized by the transverse 
condensate length $a_{\bot}$ in reducing the dimensionality of the problem from
3D to 1D \cite{olshanii}.
\bibitem{foot2} The modes may be defined in many ways, e.g. as states
of potentials obtained 
by expanding Eq.~({\protect \ref{pot}}) 
around a center of each well up to the second order (which, however, 
is a poor approximation for the Gaussian-like well which we deal
with in our problem) or as symmetric and antisymmetric eigenstates
combinations. Such states can be calculated as solutions
of the Gross-Pitaevskii equation (providing, in general,
a more accurate two-mode approximation \cite{smerzi}).
In the present paper, however, the localized wave-functions have
been calculated by neglecting the (in our case very small) non-linearity.
\bibitem{tmanew} In the time dependent calculations we have defined modes as the
two lowest eigenstates of the potential ({\protect \ref{pot}}).
It allows us to describe the system not only when the potential
({\protect \ref{pot}}) reveals double well structure but
in the full range of the potential parameter $x_0$.
\end{references}
\end{document}